\documentclass[twocolumn,showpacs,amsmath,amssymb,pra,superscriptaddress,floatfix]{revtex4}

\usepackage{amssymb}
\usepackage{graphicx}% Include figure files
\usepackage{dcolumn}% Align table columns on decimal point
\usepackage{bm}% bold math
\usepackage{psfrag}

\begin{document}

\title{Collective photon emission from symmetric states created with Rydberg atoms on a ring lattice}

\author{B. Olmos}
\email{beatriz.olmos-sanchez@nottingham.ac.uk}
\affiliation{Midlands Ultracold Atom Research Centre - MUARC, The University of Nottingham, School of Physics and Astronomy, Nottingham, United Kingdom}
\author{I. Lesanovsky}
\email{igor.lesanovsky@nottingham.ac.uk}
\affiliation{Midlands Ultracold Atom Research Centre - MUARC, The University of Nottingham, School of Physics and Astronomy, Nottingham, United Kingdom}

\date{\today}
\begin{abstract}
We discuss the creation of non-classical light from collective atomic states that are prepared in a ring-shaped lattice. These states are realized by exploiting the strong interaction between atoms in high lying energy levels - so-called Rydberg states - and yield a resource for creating excitations of the electromagnetic field that carry few photons. We characterize the properties of these photonic states showing that they are determined by the interplay between the ring geometry, the structure of the atomic resource states and the collectivity in the photon emission which is controlled by the lattice spacing. The system permits the creation of single photons with well-defined orbital angular momentum and two-photon states that are entangled in orbital angular momentum.
\end{abstract} \maketitle

\section{Introduction}
The coupling of atoms to light has been exploited since the early days of atomic physics to gain an insight into the structure of atoms and molecules. Recently, there has been a great deal of interest in the quantum interface between light and an atomic ensemble \cite{Hammerer10} and in using the coherent coupling between these two systems for quantum information processing, quantum information storage or the creation of deterministic photon sources \cite{Hau99,Fleischhauer00,Balic05,Pedersen09,Nielsen10,Gorshkov10}.
Such photon sources rely on the ability to create certain entangled quantum states in atomic ensembles which are then subsequently converted into excitations of the electromagnetic field \cite{Porras08,Scully06,Mazets07}. Ultra cold atoms provide a toolbox for the creation of such atomic states. The reason is rooted in the versatility of these systems, such as the tunability of their interactions and the advanced techniques that have been developed for their trapping and manipulation \cite{Bloch08}. In particular, the exaggerated properties of highly excited (Rydberg) atoms \cite{Gallagher94} can be exploited in order to create entangled many-particle states. In a recent work \cite{Olmos09,Olmos09-3} this was shown for a system in which the atoms are confined to a deep ring lattice. Here the collective excitations of the atomic ensemble were calculated analytically and it was shown that - due to the special geometry of the ring - these excitations possess a particularly symmetric structure.

In this paper we show how these symmetric states can be used to devise single and two photon sources. The properties of the photonic states are imposed mainly by the interplay between the particular shape of the system and the collectivity in the photon emission. We provide a - to a large extent - analytic description of the photonic states and a thorough discussion of their properties, such as their angular emission characteristics as well as their spatial correlations.
We find that the ring geometry permits the preparation of photons in a superposition state of two tunable emission directions. Furthermore, the system allows to create single photons with well-defined angular momentum and entangled photon pairs.

The paper is structured as follows: In Sec. \ref{sec:mapping} we review the general atom-light mapping scheme following Refs. \cite{Lehmberg70,Agarwal70,Porras08} with particular emphasis on the special features emerging from the ring geometry. In Sec. \ref{sec:state_preparation} we outline - along the lines of Refs. \cite{Lukin01,Olmos09,Olmos09-3} - how symmetric entangled atomic states in a ring lattice can be created by exploiting the unique properties of excited Rydberg states. These states serve as a resource for the creation of single and two photon states whose properties are thoroughly discussed in Sec. \ref{sec:single_photon} and Sec. \ref{sec:two_photon}, respectively. A conclusion and an outlook is provided in Sec. \ref{sec:conclusion_outlook}.

\section{Atom-photon mapping in a ring lattice}\label{sec:mapping}
In this section, we recapitulate the general problem of how to create photons from atomic ensembles prepared in collective quantum states. We will discuss in detail the approximations entering our calculation and illuminate peculiarities which emerge form the particularly symmetric shape of the system.
\begin{figure}
\flushleft
  \includegraphics[width=\columnwidth]{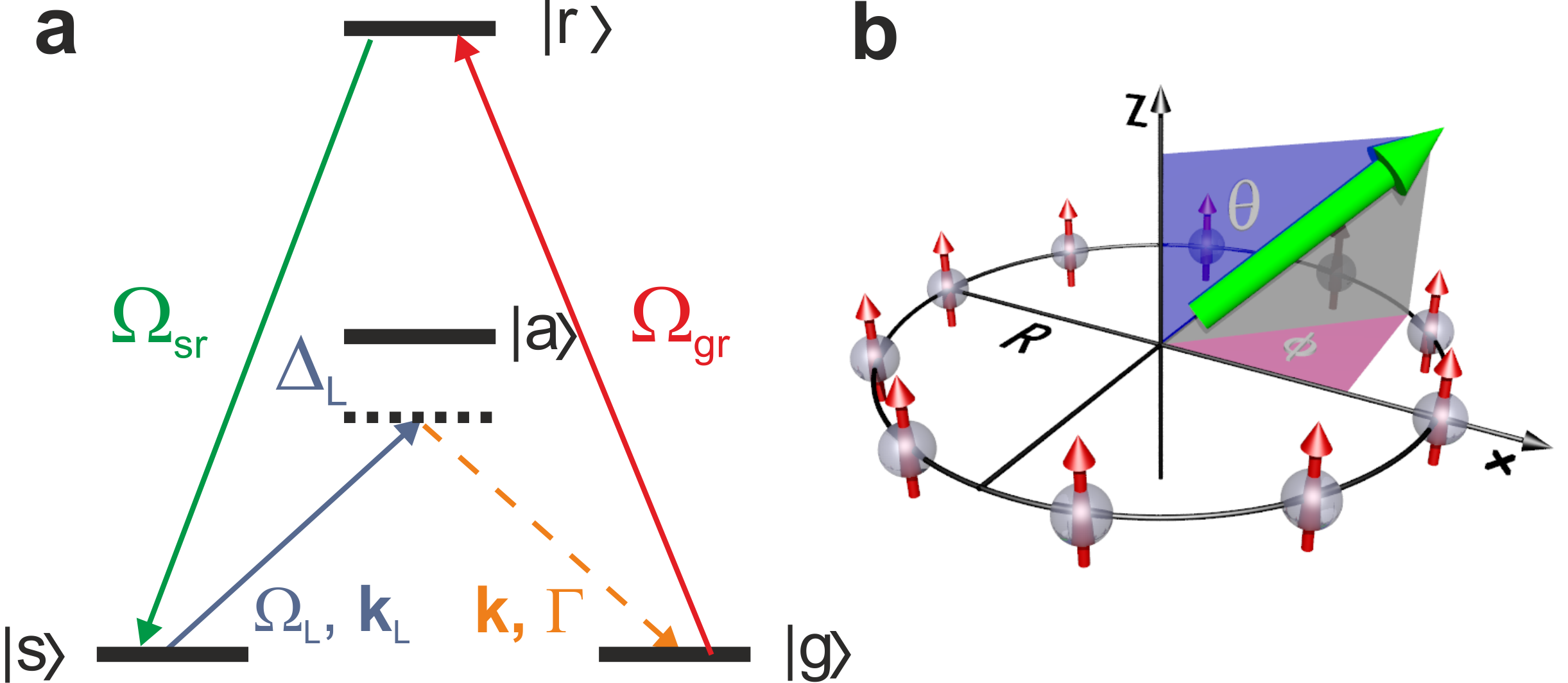}
  \caption{\textbf{a}: Internal level structure of each of the $N$ atoms of the ensemble. The excitations are stored in the two hyperfine ground states $\left|g\right>$ and $\left|s\right>$ which, together with the auxiliary level $\left|a\right>$, form a $\Lambda$-scheme: $\left|s\right>$ is coupled off-resonantly to $\left|a\right>$ and the decay from $\left|a\right>$ to $\left|g\right>$ produces emission of photons. The coupling from $\left|g\right>$ to the Rydberg state $\left|r\right>$ is used to create the initial entangled many-body atomic excitations that are subsequently mapped into the hyperfine manifold. \textbf{b}: We consider $N$ atoms confined in a ring shaped one-dimensional lattice. The transition dipoles $\mathbf{d}_{ga}$ represented by the small red arrows are aligned and perpendicular to the ring. The photons are eventually emitted from the ensemble with a certain angular distribution which is parameterized by the angles $\theta$ and $\phi$.}\label{fig:lambda}
\end{figure}

\subsection{General aspects of the atom-photon mapping}
Our setup consists of $N$ identical atoms that are spatially localized at the positions $\mathbf{r}_\alpha$. Internally, we consider three levels which form a so-called $\Lambda$-system depicted in the lower half of Fig. \ref{fig:lambda}a (note that, for the moment, we will not be concerned with the state $\left|r\right>$). The three levels involved are the two hyperfine ground states $\left|g\right>$ and $\left|s\right>$ as well as a third level $\left|a\right>$ which will be used as an intermediate state. The left 'leg' of the $\Lambda$-system is formed by a (classical) driving laser field with Rabi frequency $\Omega_\mathrm{L}$ and momentum $\mathbf{k}_\mathrm{L}$ coupling $\left|s\right>$ off-resonantly to $\left|a\right>$ with a detuning $\Delta_\mathrm{L}$. The right 'leg' is formed by the electromagnetic (quantum) field into which eventually single photons are emitted due to the decay from $\left|a\right>$ to $\left|g\right>$. In our scheme we do not consider decay from $\left|a\right>$ back to $\left|s\right>$, which can be ensured by an appropriate choice of the atomic levels.

The collective atomic excitations which will be converted into photons are stored in the two levels $\left|g\right>$ and $\left|s\right>$. These collective states have the general form
\begin{equation}\label{eqn:initial_atomic}
  \left|\Psi\right>_\mathrm{at}=\sum_{ij\dots}\Psi_{ij\dots}\sigma_{sg}^{(i)}\sigma_{sg}^{(j)} \dots \left|0\right>_\mathrm{at},
\end{equation}
with $i,j\dots=1\dots N$ and $\left|0\right>_\mathrm{at}=\prod_{k}\left|g\right>_k$ being the atomic vacuum state. The $\sigma$-operators are defined as
\begin{equation*}
  \sigma_{\alpha\beta}^{(k)}\equiv\left|\alpha\right>_k\left<\beta\right|.
\end{equation*}

Let us now show how such an atomic excitation is converted into photons: We assume that $\left|\Delta_\mathrm{L}\right|\gg\Omega_\mathrm{L}$ which allows us to adiabatically eliminate the intermediate state $\left|a\right>$. The whole problem reduces then to an ensemble of two level systems which are coupled to a multi-mode light field with Hamiltonian (in the rotating-wave approximation)
\begin{eqnarray}\label{eqn:hamiltonian}
  H&=&\sum_{\alpha=1}^N\omega_\mathrm{L}\sigma_{ss}^{(\alpha)}+ \sum_{\mathbf{q}\lambda}\omega_{\mathbf{q}}a^\dagger_{\mathbf{q}\lambda}a_{\mathbf{q}\lambda}\\\nonumber &&-\sum_{\alpha=1}^N\sum_{\mathbf{q}\lambda}K_{\mathbf{q}\lambda} \left(e^{i\left(\mathbf{q}-\mathbf{k}_\mathrm{L}\right)\cdot\mathbf{r}_\alpha}a_{\mathbf{q}\lambda}\sigma^{(\alpha)}_{sg}+ \mathrm{H.c.}\right).
\end{eqnarray}
The first and second terms of (\ref{eqn:hamiltonian}) represent the atomic energy ($\omega_\mathrm{L}=c\left|\mathbf{k}_\mathrm{L}\right|$) and the energy of the electromagnetic field, respectively. Here the operators
$a_{\mathbf{q}\lambda}$ ($a_{\mathbf{q}\lambda}^\dagger$) annihilate (create) photons with energy $\omega_\mathbf{q}=c|\mathbf{q}|$ and unit polarization vector $\mathbf{e}_{\mathbf{q}\lambda}$ ($\mathbf{q}\cdot\mathbf{e}_{\mathbf{q}\lambda}=0$). The atom-field interaction is described in the third term of the Hamiltonian with the coupling coefficients being defined as
\begin{equation*}
  K_{\mathbf{q}\lambda}=\left(\frac{\Omega_\mathrm{L}}{\Delta_\mathrm{L}}\right)\sqrt{\frac{\omega_{\mathbf{q}}}{2\epsilon_0 V}}\,\mathbf{d_{\mathrm{ga}}}\cdot\mathbf{e}_{\mathbf{q}\lambda}.
\end{equation*}
Here, $V$ is the quantization volume, $\epsilon_0$ the vacuum permitivity, and $\mathbf{d_{\mathrm{ga}}}$ the dipole matrix element of the $\left|g\right>\rightarrow\left|a\right>$ transition.

Upon switching on the classical field with Rabi frequency $\Omega_\mathrm{L}$ the time-evolution under the Hamiltonian (\ref{eqn:hamiltonian}) will convert the atomic state (\ref{eqn:initial_atomic}) into a photonic state. For times much longer than the lifetime of the intermediate state $\tau=\Gamma^{-1}$, where $\Gamma$ is its corresponding decay rate to $\left|g\right>$, this can be formulated as a direct mapping between atomic and photonic states \cite{Porras08,Scully06}. This mapping is expressed by the unitary transformation
\begin{equation}\label{eqn:mapping}
  U\sigma_{sg}^{(\alpha)}U^\dagger\stackrel{t\gg\tau}{=}\sum_{\mathbf{q}\lambda} g_{\alpha \mathbf{q}\lambda}(t)a^\dagger_{\mathbf{q}\lambda},
\end{equation}
with the time-evolution operator $U\equiv e^{-iHt}$. The coefficients appearing in this expression can be calculated according to
\begin{eqnarray}\label{eqn:g-coef-2}
  g_{\alpha \mathbf{q}\lambda}(t)&=&iK_{\mathbf{q}\lambda}\sum_{\gamma}e^{-i\left(\mathbf{q}-\mathbf{k}_\mathrm{L}\right)\cdot\mathbf{r}_\gamma}\\\nonumber
  &&\times\int_{0}^{t} e^{-i\omega_{\mathbf{q}}(t-\tau)} {}_\mathrm{at}\left<0\right|\sigma_{gs}^{(\gamma)}(\tau)\sigma_{sg}^{(\alpha)}\left|0\right>_\mathrm{at}d\tau,
\end{eqnarray}
where ${}_\mathrm{at}\left<0\right|\sigma_{gs}^{(\gamma)}(\tau)\sigma_{sg}^{(\alpha)}\left|0\right>_\mathrm{at}$ represents the atom-atom correlation function.

Our final aim is to find an explicit expression for the photonic state $\left|\Phi\right>_\mathrm{ph}$ onto which a given collective atomic state is mapped, i.e.,
\begin{equation}\label{eqn:photon_general}
  \left|\Phi\right>_\mathrm{ph}=\sum_{ij\dots}\Psi_{ij\dots} U\sigma_{sg}^{(i)} U^\dagger U\sigma_{sg}^{(j)} U^\dagger\dots\left|0\right>_\mathrm{ph},
\end{equation}
where $\left|0\right>_\mathrm{ph}$ is the photonic vacuum. Hence, we need to know the exact form of the coefficients (\ref{eqn:g-coef-2}) which is obtained from the time evolution of the correlation function, i.e., of the atomic operators $\sigma^{(\gamma)}_{gs}(\tau)$ \cite{Lehmberg70,Agarwal70}. This evolution is governed by the master equation
\begin{equation*}
  \dot{\rho}=\sum_{\alpha\beta}e^{-i\mathbf{k}_\mathrm{L}\cdot\mathbf{r}_{\alpha\beta}}J_{\alpha\beta}\left(\sigma^{(\beta)}_{gs}\rho\sigma^{(\alpha)}_{sg}- \rho\sigma^{(\alpha)}_{sg}\sigma^{(\beta)}_{gs}\right)+\mathrm{H.c.},
\end{equation*}
where the entries of the matrix $J$ are given by
\begin{equation*}
  J_{\alpha\beta}=\gamma_{\alpha\beta}+i\Omega_{\alpha\beta},
\end{equation*}
with
\begin{eqnarray*}
  \gamma_{\alpha\beta}&=&\frac{3\Gamma}{2}\left\{\left[1-3\left( \hat{d}_{ga}\cdot\hat{r}_{\alpha\beta} \right)^2\right]\left[\frac{\cos{\kappa_{\alpha\beta}}}{\kappa_{\alpha\beta}^2}- \frac{\sin{\kappa_{\alpha\beta}}}{\kappa_{\alpha\beta}^3}\right]\right.\\
  &&\left.+\left[1-\left( \hat{d}_{ga}\cdot\hat{r}_{\alpha\beta} \right)^2\right]\frac{\sin{\kappa_{\alpha\beta}}}{\kappa_{\alpha\beta}}\right\}\\
  \Omega_{\alpha\beta}&=&\frac{3\Gamma}{2}\left\{\left[1-3\left( \hat{d}_{ga}\cdot\hat{r}_{\alpha\beta} \right)^2\right]\left[\frac{\sin{\kappa_{\alpha\beta}}}{\kappa_{\alpha\beta}^2}+ \frac{\cos{\kappa_{\alpha\beta}}}{\kappa_{\alpha\beta}^3}\right]\right.\\
  &&\left.-\left[1-\left( \hat{d}_{ga}\cdot\hat{r}_{\alpha\beta} \right)^2\right]\frac{\cos{\kappa_{\alpha\beta}}}{\kappa_{\alpha\beta}}\right\}.
\end{eqnarray*}
In these expressions, $\Gamma=\left(\frac{\Omega_\mathrm{L}}{\Delta_\mathrm{L}}\right)^2\frac{d_{ga}^2k_\mathrm{L}^3}{6\pi\epsilon_0}$ is the single-atom decay rate and $\kappa_{\alpha\beta}\equiv k_\mathrm{L}\left|\mathbf{r}_{\alpha\beta}\right|$, with $\mathbf{r}_{\alpha\beta}=\mathbf{r}_\alpha-\mathbf{r}_{\beta}$ and $k_\mathrm{L}=\left|\mathbf{k}_\mathrm{L}\right|$.

Some features of the photon emission can be already inferred from the eigenvalues of $J$. Its largest eigenvalue $\Gamma_\mathrm{col}$ defines the 'degree of collectivity' in the photon emission which in general depends on the interplay between two parameters: the average interparticle distance, $a$, and the wavelength, $\lambda_\mathrm{L}$. If the wavelength of the laser is much larger than the separation between the atoms, the whole ensemble couples to the light field as a single degree of freedom. Hence, in this case the degree of collectivity is large, i.e. $\Gamma_\mathrm{col}\gg\Gamma$.
In the opposite regime, i.e. $a\gg\lambda_\mathrm{L}$, the atoms couple independently to the laser and, hence, $\Gamma_\mathrm{col}=\Gamma$.
Throughout this paper, we will focus mainly in the more interesting intermediate regime $a\sim\lambda_\mathrm{L}$. There, a non-negligible degree of collectivity is present while the specific spatial configuration of the atoms give rise to directionality and other features that we will discuss in the next sections.

We proceed by employing a final approximation which relies on the fact that we are particularly interested in atomic states in which the number of atoms in the state $\left|s\right>$ is very small in comparison to the total number of atoms $N$. In this restricted subspace - denoted by $\langle\!\!\!\langle...\rangle\!\!\!\rangle$ - the commutation relations between the spin operators can be approximated by
\begin{equation}\label{eqn:bosonic}
  \langle\!\!\!\langle\left[\sigma^{(\alpha)}_{gs},\sigma^{(\beta)}_{sg}\right]\rangle\!\!\!\rangle
  =\left(1-2\langle\!\!\!\langle\sigma^{(\alpha)}_{ss}\rangle\!\!\!\rangle\right)\delta_{\alpha\beta} \approx\delta_{\alpha\beta},
\end{equation}
i.e., the $\sigma$-operators obey a {\it bosonic algebra}. This is essentially the Holstein-Primakoff approximation \cite{Holstein40}. Dropping the notation $\langle\!\!\!\langle...\rangle\!\!\!\rangle$ in the following, the time evolution of the expectation value of $\sigma^{(\gamma)}_{gs}(\tau)$ can be written in the closed form:
\begin{equation*}
  \frac{d\left<\sigma^{(\gamma)}_{gs}(t)\right>}{dt}=-\sum_{\beta}e^{-i\mathbf{k}_\mathrm{L}\cdot\mathbf{r}_{\gamma\beta}} J_{\gamma\beta}\left<\sigma^{(\beta)}_{gs}(t)\right>.
\end{equation*}
Using the quantum regression theorem \cite{Gardiner04,Breuer06}, we obtain for the time-evolution of the correlation function
\begin{equation*}
  \frac{d\left<\sigma^{(\gamma)}_{gs}(\tau)\sigma^{(\alpha)}_{sg}\right>}{d\tau}= -\sum_{\beta}e^{-i\mathbf{k}_\mathrm{L}\cdot\mathbf{r}_{\gamma\beta}} J_{\gamma\beta}\left<\sigma^{(\beta)}_{gs}(\tau)\sigma^{(\alpha)}_{sg}\right>.
\end{equation*}
In order to solve this equation of motion, we introduce the eigenfunctions and eigenvalues of the operator $J$
\begin{eqnarray}\label{eqn:diag_J}
  J_{\gamma\beta}=\sum_{mn}{\cal M}_{\gamma n}D_n\delta_{nm}{\cal M}^{-1}_{m\beta},
\end{eqnarray}
such that the desired expectation value of the correlations yields
\begin{eqnarray*}
  \left<\sigma^{(\gamma)}_{gs}(\tau)\sigma^{(\alpha)}_{sg}\right>=e^{i\mathbf{k}_\mathrm{L}\cdot\mathbf{r}_{\alpha\gamma}} \sum_{k}{\cal M}_{\gamma k}e^{-D_k\tau}{\cal M}^{-1}_{k\alpha}.
\end{eqnarray*}
Introducing this result into equation (\ref{eqn:g-coef-2}), we eventually obtain (in the limit of $t\gg 1/\Gamma$),
\begin{equation}\label{eqn:g}
  g_{\alpha \mathbf{q}\lambda}(t)=-iK_{\mathbf{q}\lambda}e^{-i\left(\omega_{\mathbf{q}}t-\mathbf{k}_\mathrm{L}\cdot\mathbf{r}_{\alpha}\right)}\sum_{\gamma k}e^{-i\mathbf{q}\cdot\mathbf{r}_\gamma}\frac{{\cal M}_{\gamma k}{\cal M}^{-1}_{k\alpha}}{i\omega_{\mathbf{q}}-D_k}.
\end{equation}
These coefficients contain all the information of the mapping (\ref{eqn:mapping}) and put us in a position to calculate the photonic state (\ref{eqn:photon_general}) that is created by an atomic excitation of the form (\ref{eqn:initial_atomic}).

\subsection{The ring lattice configuration}
After these general considerations let us now focus on the particular ring structure of our system. The $N$ atoms are placed on a ring lattice (one atom per site) with interatomic spacing $a$. We consider that the ring lies in the $xy$-plane. This setup is shown in Fig. \ref{fig:lambda}b. The position of each atom is thus given by
\begin{equation*}
  \mathbf{r}_\alpha=R\left(\cos{\phi_\alpha},\sin{\phi_\alpha},0\right),
\end{equation*}
with $\phi_\alpha=\frac{2\pi}{N}(\alpha-1)$, for $\alpha=1\dots N$. Here, $R$ is the radius of the ring and it is given approximately (for large number of sites $N\gg1$) by $R\approx\frac{aN}{2\pi}$.

We assume furthermore that the dipoles of the transition $\left|g\right>\rightarrow\left|a\right>$ are oriented perpendicular to the ring, i.e. $\hat{d}_{ga}\parallel \hat{z}$. This leads to a particularly simple appearance of the the operator $J$ which governs the evolution of the atomic operators. Its matrix representation becomes a circulant matrix, that is, each row contains the elements of the previous one shifted cyclically one place to the right. In addition, as a consequence of the periodic boundary conditions, $J$ is symmetric. This matrix can be diagonalized analytically \cite{Tee05}, and its eigenfunctions and eigenvalues are
\begin{equation}\label{eqn:eigen}
  {\cal M}_{\gamma k}=\frac{e^{i\phi_{k}(\gamma-1)}}{\sqrt{N}},\quad D_k=\sum_{n=1}^{N}J_{1n}e^{i\phi_k(n-1)},
\end{equation}
for $k=1\dots N$.
This result allows a quasi-analytical treatment of the photon emission by the collective atomic excitations stored in the ring lattice.

\section{State preparation} \label{sec:state_preparation}
Having now understood how to map atomic excitations into photonic states we will briefly outline in this section which many-particle states are accessible in the ring lattice, and hence which states can serve as a resource for emitted photons.

We follow the schemes discussed in Refs. \cite{Lukin01,Olmos09,Olmos09-3} which rely on the properties of atoms excited to Rydberg states. To this end we introduce for each atom the Rydberg ns-state $\left|r\right>$ which is coupled to the electronic ground state $\left|g\right>$ by means of a laser with Rabi frequency $\Omega_{gr}$ (see Fig. \ref{fig:lambda}a). When the atoms are in the excited state, they interact via the van-der-Waals interaction $V_\mathrm{vdW}(\mathbf{r})=C_6\times\left|\mathbf{r}\right|^{-6}$, where $\mathbf{r}$ is the separation between the atoms and $C_6$ is the van-der-Waals coefficient \cite{Marinescu95,Singer05}. In the case of Rydberg states, this interaction can be very strong even over large spatial separations. This gives rise to the so-called 'Rydberg blockade', that inhibits the simultaneous excitation of more than one Rydberg atom inside a sphere with \emph{blockade radius} $r_\mathrm{b}\sim \left[C_6\,\Omega_{gr}^{-1}\right]^{1/6}$ \cite{Jaksch00,Lukin01}. The size of the blockade radius relative to the interparticle distance $a$ will crucially determine the evolution dynamics of the system under the action of the laser coupling from $\left|g\right>$ to $\left|r\right>$.

\begin{figure}
\center
\includegraphics[width=\columnwidth]{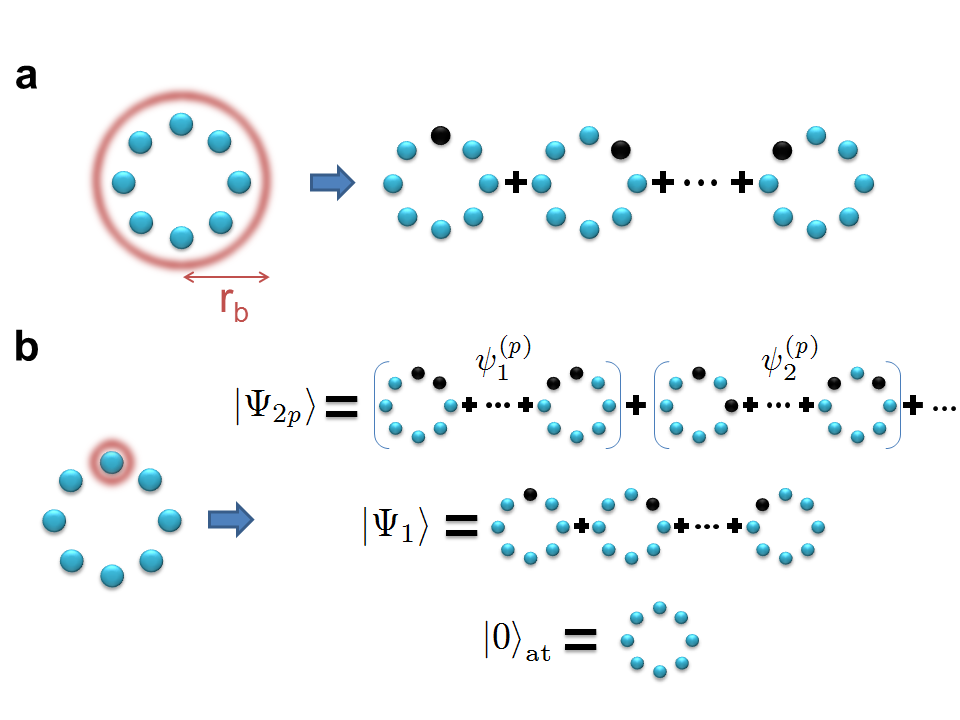}
\caption{\textbf{a}: When the blockade radius, represented by the blurred red circle, is larger than the radius of the ring, the laser excites a symmetric superposition of all possible single atomic excitations. \textbf{b}: When the laser driving is much stronger than the interaction, i.e., $r_\mathrm{b}\ll a$, one can write exactly the eigenstates of the system. On the right column, an sketch of the ground state and first excited states is depicted.} \label{fig:blockade}
\end{figure}

We will make a distinction between the cases in which $r_\mathrm{b}$ is (i) so large that it encompasses the entire lattice and (ii) smaller than the interparticle separation. We discuss the corresponding schemes for the creation of entangled atomic states for these two cases in the following:

\textbf{(i) - Large blockade radius.}  This case is depicted in Fig. \ref{fig:blockade}a. Here the entire lattice is blockaded and the laser coupling $\left|g\right>$ to $\left|r\right>$ can only excite a single atom to the Rydberg state. As a consequence, the laser effectively couples the two collective states $\left|0\right>_\mathrm{at}\equiv\prod_k\left|g\right>_k$ and $\left|R\right>\equiv1/\sqrt{N}\sum_{k}\sigma_{rg}^{(k)}\left|0\right>_\mathrm{at}$ with a collective Rabi frequency $\Omega=\sqrt{N}\Omega_{gr}$ \cite{Lukin01}.
Hence, as a result of a pulse of duration $\tau_{gr}=\pi/\Omega$ a symmetric superposition of all possible single atomic excitations is achieved. Subsequently the excited atomic states are mapped onto the stable hyperfine state $\left|s\right>$ (see Fig. \ref{fig:lambda}a) such that the spin wave state
\begin{equation}\label{eqn:spin_wave}
  \left|\Psi_1\right>=\frac{1}{\sqrt{N}}\sum_{\alpha=1}^N\sigma^{(\alpha)}_{sg}\left|0\right>_\mathrm{at}
\end{equation}
is reached.
Note that we have considered here that all atoms are located in a plane with constant phase equal to zero, i.e., the momenta of the involved lasers are perpendicular to the ring.

\textbf{(ii) - Small blockade radius.} The previous and other entangled atomic states can also be achieved in the regime of a very weak blockade which is sketched in Fig. \ref{fig:blockade}b. The corresponding scheme is described in \cite{Olmos09-3,Olmos10-1,Olmos10-2} and we only outline the results here: When the single atom Rabi frequency $\Omega_{gr}$ is much larger than the interaction between neighboring Rydberg atoms, the Hamiltonian of this system becomes analytically solvable. One can write the ground state and first excited many-particle states in terms of the single atom states $\left|\pm\right>_k=\frac{1}{\sqrt{2}}\left[\left|g\right>_k\pm\left|r\right>_k\right]$. In particular, the ground state of the Hamiltonian is given by the product state $\left|G\right>=\prod_{k}\left|-\right>_k$.
Due to the symmetry of the Hamiltonian, only very specific - so-called fully symmetric - states are accessible from the ground state. The first excited fully symmetric state is given by
\begin{eqnarray*}
  \left|1\right>=\frac{1}{\sqrt{N}}\sum_{k=1}^N\sigma_+^{(k)}\left|G\right>,
\end{eqnarray*}
with $\sigma_\pm^{(k)}\left|\mp\right>_k=\left|\pm\right>_k$. Here only one of the atoms in the ring is in the state $\left|+\right>$, but this excitation is delocalized over the entire ring. One can excite also entangled fully symmetric states which carry two excitations. These are
\begin{equation*}
  \left|2_p\right>=\frac{1}{N}\sum_{kk'}\sin{\left[\frac{2\pi}{N}(p-1/2)|k-k'|\right]}\sigma_+^{(k)}\sigma_+^{(k')}\left|G\right>,
\end{equation*}
with $p=1\dots\lfloor N/2\rfloor$. Note that these states are superpositions of pairs of excitations that travel in the ring with opposite momentum. Let us point out that the selective excitation of the above-described states is experimentally possible by means of variations in the Rabi frequency and detuning of the laser, as it is described in Refs. \cite{Olmos09-3,Olmos10-1,Olmos10-2}.

Once the desired entangled states are achieved, one maps the excited states which are encoded in the superpositions $\left|+\right>$ and $\left|-\right>$ to the stable states of the atomic hyperfine groundstate manifold.
This is done by means of a sequence of two resonant laser pulses (see Ref. \cite{Olmos10-2} for a detailed explanation) that perform the mappings $\left|-\right>_k\rightarrow \left|g\right>_k$ and $\left|+\right>_k\rightarrow i\left|s\right>_k$. This brings the many-particle states described before into the states $\left|0\right>_{\mathrm{at}}$, (\ref{eqn:spin_wave}) and
\begin{equation}\label{eqn:doubly-excited}
  \left|\Psi_{2_p}\right>=\sum_{kk'}\psi_{kk'}^{(p)}\sigma^{(k)}_{sg}\sigma^{(k')}_{sg}\left|0\right>_\mathrm{at},
\end{equation}
with $\psi_{kk'}^{(p)}=1/N \sin{\left[\frac{2\pi}{N}(p-1/2)|k-k'|\right]}$, respectively (see Fig. \ref{fig:blockade}b).

We can now use the atom-light mapping presented in Sec. \ref{sec:mapping} to calculate the photonic states that are created from the collective atomic excitations (\ref{eqn:spin_wave}) and (\ref{eqn:doubly-excited}). The spin wave (\ref{eqn:spin_wave}) produces the single-photon state
\begin{equation}\label{eqn:single-photon}
  \left|\Phi_1\right>=\frac{1}{\sqrt{N}}\sum_{\mathbf{q}\lambda}\sum_{\alpha=1}^N g_{\alpha\mathbf{q}\lambda}(t) a^\dag_{\mathbf{q}\lambda}\left|0\right>_{\mathrm{ph}}.
\end{equation}
The doubly excited states (\ref{eqn:doubly-excited}) which are characterized by the label $p$ convert to two-photon states which possess the form
\begin{equation}\label{eqn:double-photon}
  \left|\Phi_{2_p}\right>=\sum_{\mathbf{q}\lambda\mathbf{q}'\lambda'}\sum_{kk'} \psi_{kk'}^{(p)}g_{k\mathbf{q}\lambda}(t)g_{k'\mathbf{q}'\lambda'}(t)a_{\mathbf{q}\lambda}^\dag a_{\mathbf{q}'\lambda'}^\dag\left|0\right>_\mathrm{ph}.
\end{equation}
The coefficients $g_{\alpha\mathbf{q}\lambda}(t)$ are given by Eq. (\ref{eqn:g}).

\section{Creation of single photons} \label{sec:single_photon}
Now we perform a detailed analysis of the photonic excitations focussing at first on the single photon state (\ref{eqn:single-photon}). The properties of the emitted photon will depend on the degree of collectivity and the orientation of the laser momentum $\mathbf{k}_\mathrm{L}$ with respect to the ring plane.

An important quantity for the characterization of the photonic state is the angular intensity distribution, i.e. the average photon number per solid angle. It is defined through
\begin{eqnarray}\label{eqn:intensity_general}
  I(\theta,\phi)=\frac{V}{(2\pi c)^3}\int_0^\infty \sum_{\nu}\left<n_{\mathbf{q}\nu}\right>\omega_\mathbf{q}^2d\omega_\mathbf{q},
\end{eqnarray}
with the number operator being $n_{\mathbf{q}\nu}=a^\dag_{\mathbf{q}\nu}a_{\mathbf{q}\nu}$ and the angles $\theta$ and $\phi$ determining the direction of the emission (see Fig. \ref{fig:lambda}b).
In the particular case of our single photon state, one can show that the angular intensity yields
\begin{eqnarray*}
  I(\theta,\phi)&=&\frac{3\Gamma\sin^2{\theta}}{4\pi N^3}\sum_{m,n=1}^N \frac{B_n(\theta_\mathrm{L},\phi_\mathrm{L})B^*_m(\theta_\mathrm{L},\phi_\mathrm{L})}{D_m+D_n^*}\\
  &&\times B^*_n(\theta,\phi)B_m(\theta,\phi),
\end{eqnarray*}
where the $D_n$ are given in (\ref{eqn:eigen}), the function $B_n(\theta,\phi)$ is
\begin{equation}\label{eqn:B}
  B_n(\theta,\phi)=\sum_{\gamma=1}^Ne^{-ik_\mathrm{L}R\,\hat{\mathbf{q}}\cdot\hat{\mathbf{r}}_\gamma}e^{i\phi_{\gamma}(n-1)},
\end{equation}
and $\left(\theta_\mathrm{L},\phi_\mathrm{L}\right)$ characterize the direction of the momentum vector $\mathbf{k}_\mathrm{L}$ of the incident laser in spherical coordinates.

\begin{figure}
\center
\includegraphics[width=7cm]{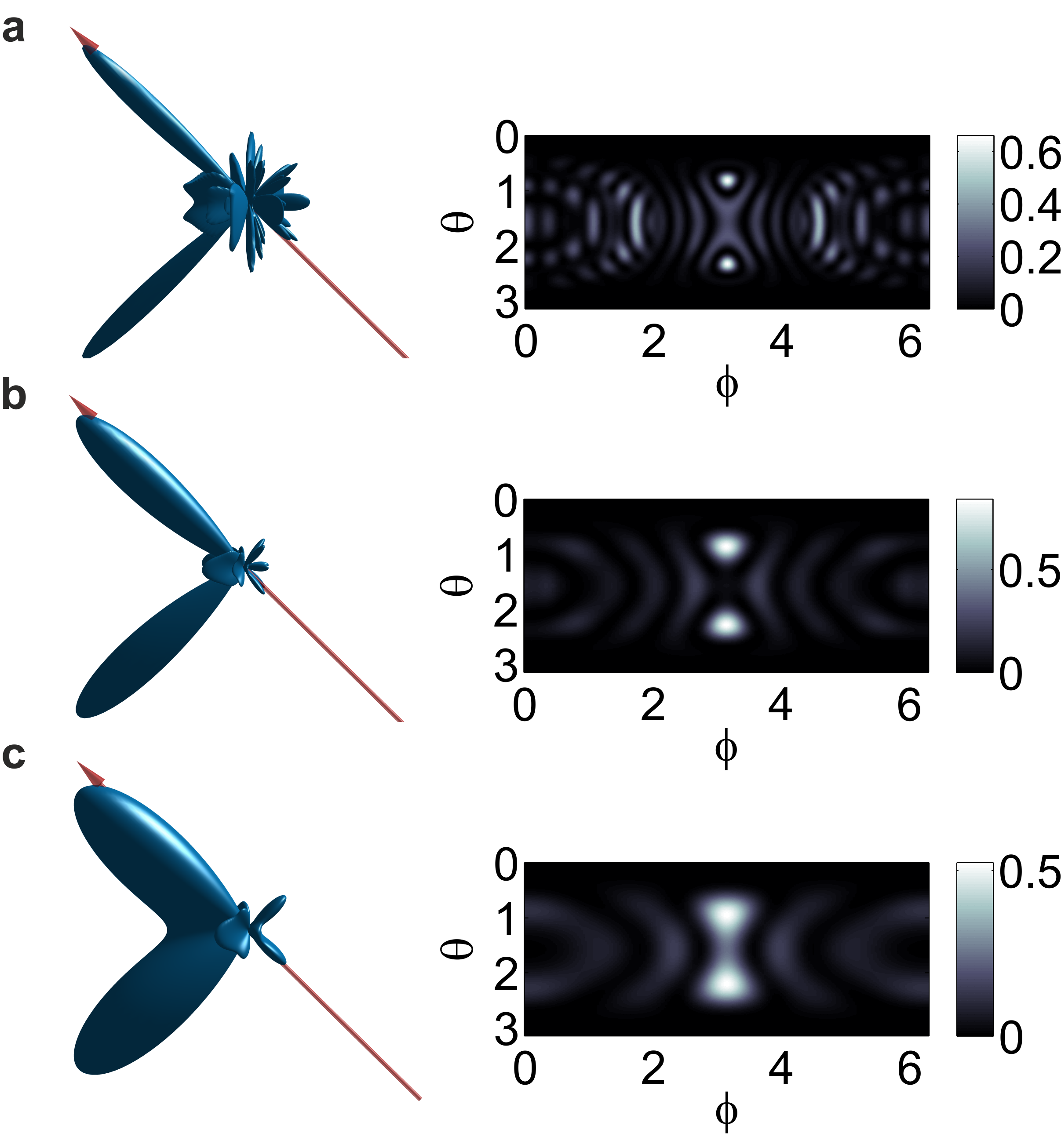}
\caption{Angular intensity distribution for a ring of $N=15$ sites and \textbf{a}: $a=\lambda_\mathrm{L}$, \textbf{b}: $a=\lambda_\mathrm{L}/2$ and \textbf{c}: $a=\lambda_\mathrm{L}/3$. The red arrow indicates the direction of $\mathbf{k}_\mathrm{L}$ which is chosen to form an angle of $45$ degrees with the vertical ($z$-)axis at an azimuthal angle $\phi_\mathrm{L}=\pi$. The structure of the intensity profile is severely influenced by the collectivity of coupling of the atoms to the radiation field. This collectivity is determined by the ratio of the lattice spacing and the laser momentum and increases from $\textbf{a}$ to $\textbf{c}$. The density plots in the right column show the same data as the three-dimensional plots.} \label{fig:intensity_45}
\end{figure}
We will first discuss the case in which $\mathbf{k}_\mathrm{L}\nparallel \hat{\mathbf{z}}$, i.e. the laser momentum is not perpendicular to the plane of the ring. In this situation we observe that, for ring radii obeying $R\gtrsim \lambda_\mathrm{L}$, two peaks dominate the angular distribution of the emitted photon as shown in Fig. \ref{fig:intensity_45}. One of these peaks follows the direction of the incident laser (indicated by the red arrow). The second peak is just the corresponding mirror image.

The origin of this distribution can be easily understood in the limit $R\gg\lambda_\mathrm{L}$, in which the eigenvalues $D_n$ are independent of $n$ and all equal to $\Gamma$ ($\Gamma_\mathrm{col}=\Gamma$), i.e. in the absence of collectivity. The angular distribution of the emitted photon in this limit becomes
\begin{eqnarray*}
  I_{R\gg\lambda_\mathrm{L}}(\theta,\phi)\approx\frac{3\sin^2{\theta}}{8\pi N}\left|\sum_{\gamma=1}^Ne^{ik_\mathrm{L}R\,\left(\hat{\mathbf{q}}-\hat{\mathbf{k}}_\mathrm{L}\right)\cdot\hat{\mathbf{r}}_\gamma}\right|^2.
\end{eqnarray*}
It is known that in the case of a gas confined to a one dimensional linear lattice or in a disordered cigar-shaped cloud \cite{Pedersen09,Saffman02,Nielsen10}, this equation gives rise to only one dominant intensity maximum occurring in the same direction of the laser momentum, i.e., when $\hat{\mathbf{q}}_\mathrm{max}=\hat{\mathbf{k}}_\mathrm{L}$.
In the case of the ring this situation is different. Here the geometry leads to a photon emission into two dominant directions which are given by $\phi_\mathrm{max}=\phi_\mathrm{L}$ and $\sin{\theta_\mathrm{max}}=\sin{\theta_\mathrm{L}}$. In particular, when $\theta_\mathrm{L}=\pi/4$, the two peaks of the intensity are perpendicular to each other which corresponds to the example shown in Fig. \ref{fig:intensity_45}. In each panel of the figure we show the intensity profile for different values of $a/\lambda_\mathrm{L}$ (determining the degree of collectivity) and $N=15$ atoms. One clearly sees that with increasing $a/\lambda_\mathrm{L}$ the main peaks get sharper but at the same time the number of smaller peaks increases as the ensemble ceases behaving collectively in the emission process.

\begin{figure}
\center
\includegraphics[width=7cm]{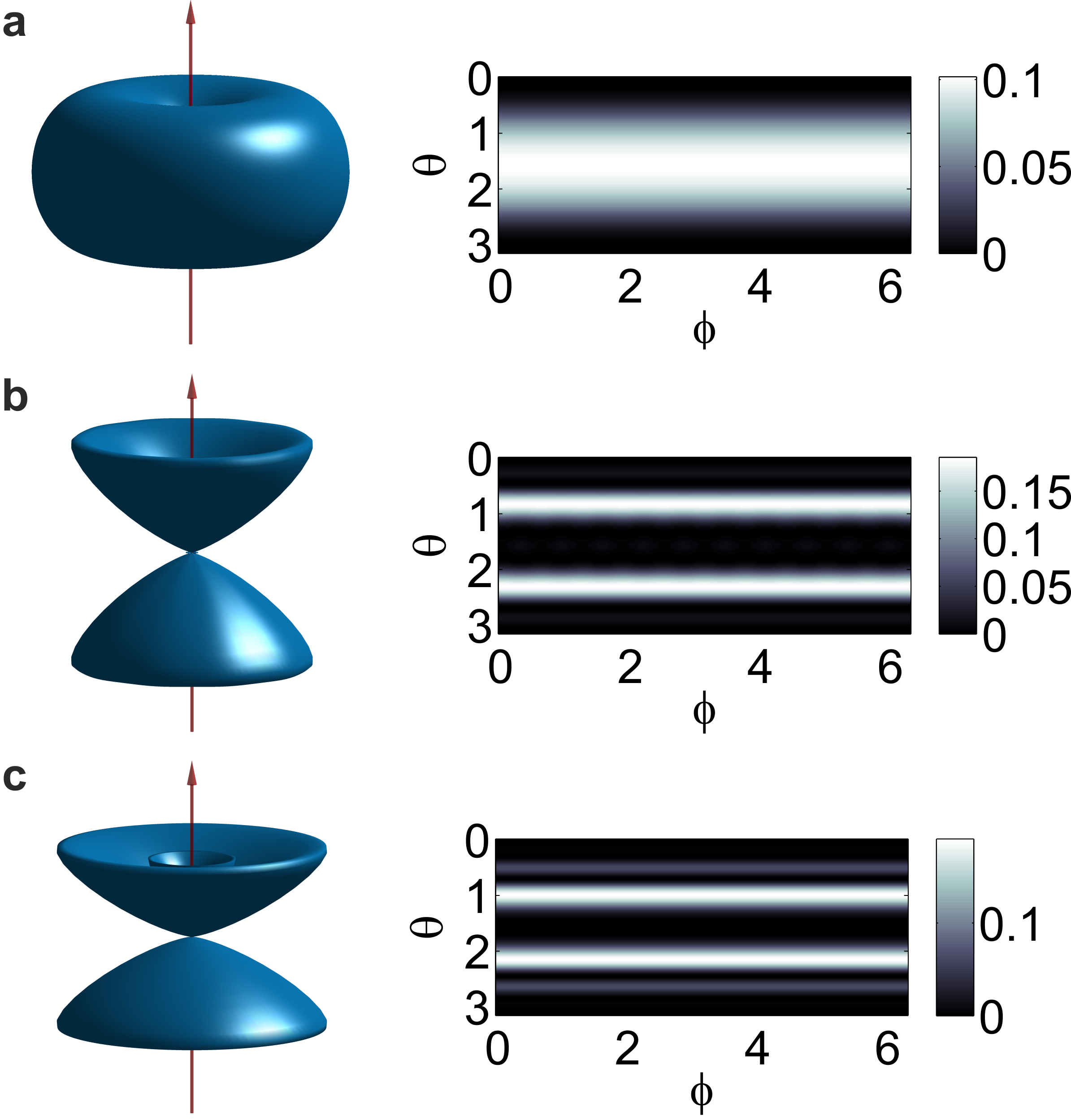}
\caption{Angular intensity distribution for several values of $a\lesssim\lambda_\mathrm{L}$: \textbf{a}: a ring of $N=10$ sites and $a=\lambda_\mathrm{L}/10$, \textbf{b}: $N=10$ and $a=0.56\lambda_\mathrm{L}$ and \textbf{c}: $N=20$ and $a=0.43\lambda_\mathrm{L}$. The density plots in the right column show the same data as the three-dimensional plots. The red arrow indicates the direction of $\mathbf{k}_\mathrm{L}$ which is chosen to be parallel to the $z$-axis. Here no dependence on the azimuthal angle is visible.} \label{fig:intensity_z}
\end{figure}
We will now turn to a particularly symmetric case, namely $\mathbf{k}_\mathrm{L}\parallel \hat{\mathbf{z}}$. Here the laser is irradiated exactly perpendicular to the ring plane. In this highly symmetric situation, the angular intensity distribution of the emitted photon takes on a particularly simple form:
\begin{equation*}
  I_{0}(\theta,\phi)=\frac{3\Gamma}{4\pi N}\frac{\sin^2{\theta}}{D_1+D_1^*}\left|\sum_{\gamma=1}^Ne^{-ik_\mathrm{L}R\, \hat{\mathbf{q}}\cdot\hat{\mathbf{r}}_{\gamma}}\right|^2.
\end{equation*}
with $D_1+D_1^*=2\sum_{n}\gamma_{1n}$ where
\begin{equation*}
  \gamma_{1n}=\frac{3\Gamma}{2}\left[\frac{\cos{\kappa_{1n}}}{\kappa_{1n}^2}- \frac{\sin{\kappa_{1n}}}{\kappa_{1n}^3}+\frac{\sin{\kappa_{1n}}}{\kappa_{1n}}\right],
\end{equation*}
and $\kappa_{1n}=k_\mathrm{L}|\mathbf{r}_{1n}|$. For sufficiently small interparticle separation $a\lesssim \lambda_\mathrm{L}$, the intensity is well approximated by
\begin{equation*}
  I_{0}(\theta,\phi)\approx\frac{3\Gamma N}{4\pi}\frac{\sin^2{\theta}}{D_1+D_1^*}J_0^2(k_\mathrm{L}R\sin{\theta}),
\end{equation*}
where $J_0(x)$ represent the zero-th order Bessel function of the first kind.
Note that, even for a finite number of atoms, this expression is independent of the azimuthal angle $\phi$. Hence, the orbital angular momentum of the emitted photon is in this case equal to zero. This is again a manifestation of the non-negligible collectivity in the photon emission. Conversely, for large values of $a\gg\lambda_\mathrm{L}$, the atoms can be approximately regarded as independent so that they are coupled individually to the radiation field. This produces strong azimuthal modulations of the intensity profile.

In Fig. \ref{fig:intensity_z} we show the angular intensity distribution obtained for three different ratios $a/\lambda_\mathrm{L}$, which are all chosen such that no azimuthal variation of $I_0(\theta,\phi)$ is present. This clearly shows that, by changing the lattice spacing, the emission characteristics of the photon can be significantly altered. For $N=10$ and $a=\lambda_\mathrm{L}/10$ (Fig. \ref{fig:intensity_z}a) the atoms are so close together that the atomic excitation (spin wave) acts as a single degree of freedom that couples to the radiation field \cite{Dicke54,Lehmberg70}. This results in an almost spherical intensity profile which is modulated by the dipole radiation pattern. In the second case (Fig. \ref{fig:intensity_z}b), with $N=10$ and $a=0.56\lambda_\mathrm{L}$, the photon emission is strongly peaked in a cone along the polar angle $\theta_{\mathrm{max}}\approx \pi/4$. By tuning the parameters $a$ and $N$ for a fixed wavelength, it is possible to shift the position of this peak and even to create a second concentric emission cone. This is seen in Fig. \ref{fig:intensity_z}c, where for $N=20$ and $a=0.43\lambda_\mathrm{L}$ the two maxima are located at $\theta_{\mathrm{max}}\approx 1$ and $0.5$, respectively.

\begin{figure}
\center
\includegraphics[width=7cm]{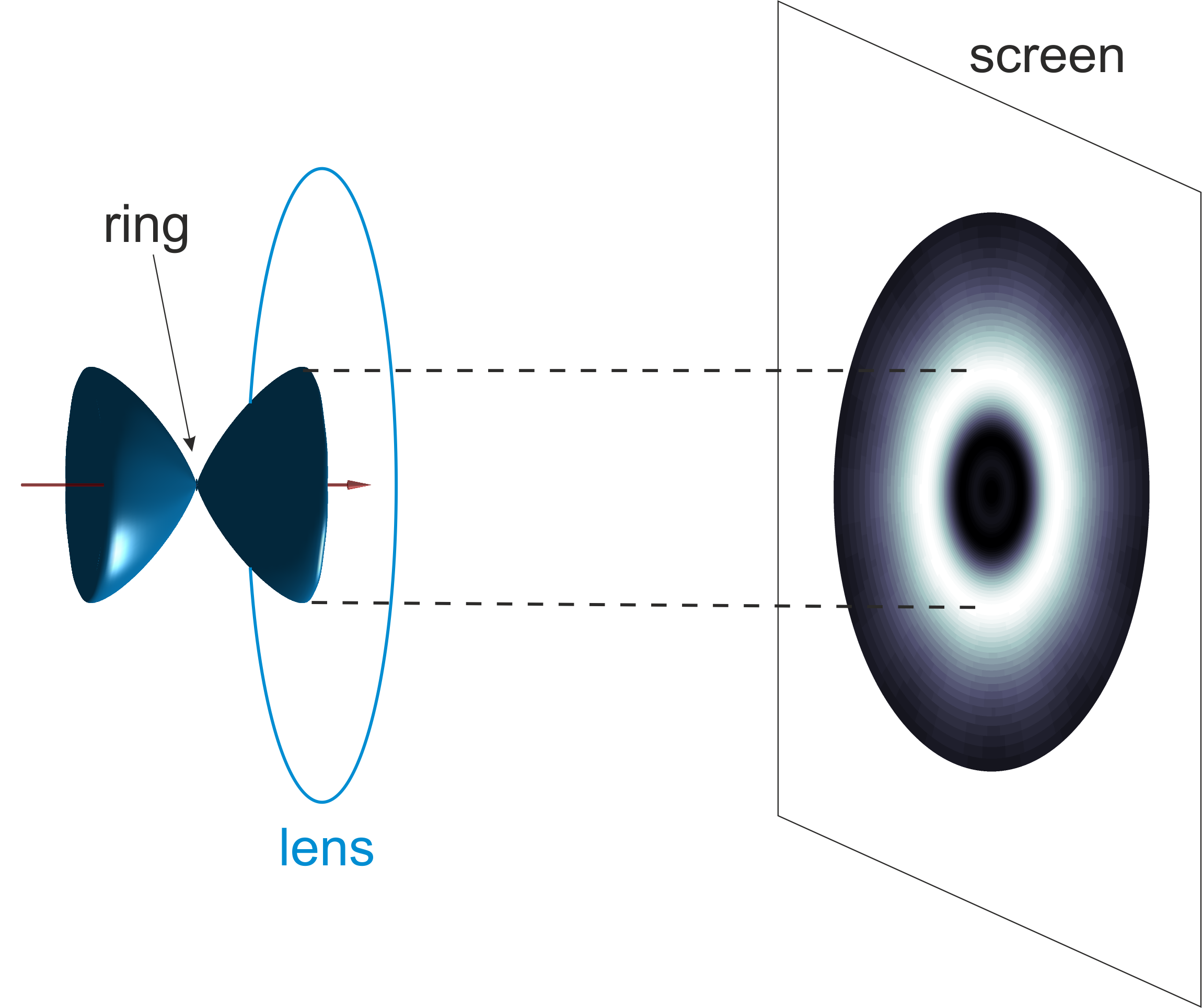}
\caption{Sketch of the projection of the hollow photon with well-defined angular momentum for $N=10$ and $a=0.56\lambda_\mathrm{L}$.} \label{fig:hollow}
\end{figure}
Due to the form of the emission, one could think of focusing the cone by means of a lens, which would give rise to a hollow photon with defined zero orbital angular momentum as depicted in  Fig. \ref{fig:hollow}. Moreover, one can find other parameter sets for which of the emission occurs mostly into the ring plane.

\section{Creation of photon pairs} \label{sec:two_photon}
Let us turn to the analysis the photon pair states that can be created using the atomic states $\left|\Psi_{2_p}\right>$ as resource. In this case, the photonic states are given by Eq. (\ref{eqn:double-photon}) and also labeled by $p$. We will show that  states with different values of $p$ possess hugely different properties.

\begin{figure}
\center
\includegraphics[width=\columnwidth]{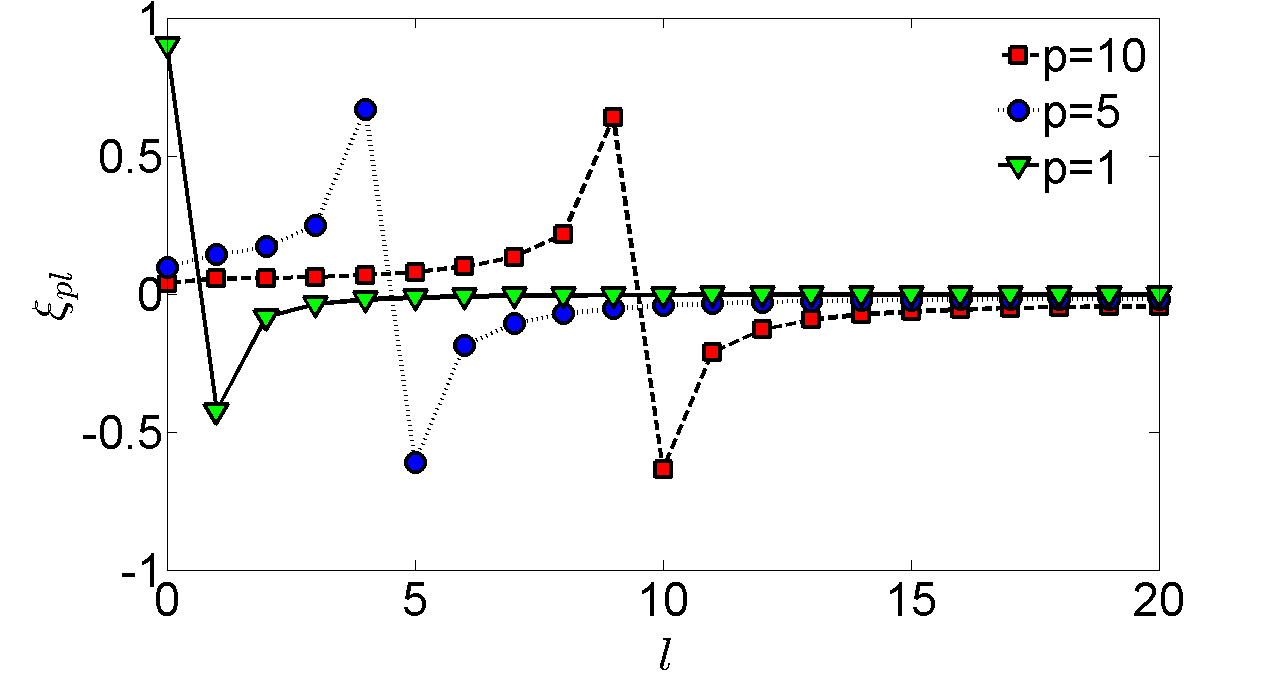}
\caption{Overlap $\xi_{pl}=\left<\Xi_l\mid\Psi_{2p}\right>$ for $N=40$. In general, the only non-negligible coefficients are the ones corresponding to $l=p$ and $l=p-1$ (note the $p=10$ and $p=5$ cases). In the case of $p=1$, the largest element is given by $l=0$. Note that all the coefficients are real.} \label{fig:density}
\end{figure}

In order to understand these differences, it is convenient to decompose the original atomic excitations (\ref{eqn:doubly-excited}) into the atomic modes
\begin{eqnarray*}
  \left|\Xi_{l}\right>&=&\frac{1}{\sqrt{1+\delta_{l0}}}\left(\frac{1}{\sqrt{N}}\sum_{k} e^{il\phi_k}\sigma^{(k)}_{sg}\right)\\ &&\times\left(\frac{1}{\sqrt{N}}\sum_{k'} e^{-il\phi_{k'}}\sigma^{(k')}_{sg}\right)\left|0\right>_\mathrm{at},
\end{eqnarray*}
with $l=0\dots N/2$. These are doubly excited states formed by two spin wave single excitations with opposite angular momentum $l$ and $-l$, respectively. In order to see that this decomposition is sensible, we calculate the overlap between the state (\ref{eqn:doubly-excited}) and the above modes: $\xi_{pl}=\left<\Xi_l\mid\Psi_{2p}\right>$. The result is shown in Fig. \ref{fig:density} for $N=40$ and three different values of $p$. We observe that, for a general value of $p$, the doubly excited states $\left|\Psi_{2_p}\right>$ can be approximately written as the entangled state $\left|\Psi_{2_p}\right>\approx\frac{1}{\sqrt{2}}\left[\left|\Xi_{p-1}\right>-\left|\Xi_{p}\right>\right]$. This can be observed in Fig. \ref{fig:density} for the values $p=5$ and $p=10$. The exception is the particular case of $p=1$ which is also shown in Fig. \ref{fig:density}. Here one observes that the state $\left|\Psi_{2_1}\right>$ can be well approximated as the zero angular momentum mode $\left|\Xi_0\right>$ (the corresponding overlap yields $\left|\xi_{10}\right|^2\approx 0.8$),
\begin{equation*}
  \left|\Psi_{2_1}\right>\approx\left|\Xi_0\right>=\frac{1}{\sqrt{2}} \left(\frac{1}{\sqrt{N}}\sum_{k=1}^N\sigma^{(k)}_{sg}\right)^2\left|0\right>_\mathrm{at},
\end{equation*}
i.e., an unentangled product state of two identical single atomic excitations. The mapping of this atomic state into a photonic one results in the emission of two identical photons.
This can be corroborated by calculating the angular distribution of the photons which is defined through (\ref{eqn:intensity_general})
\begin{eqnarray*}
  I(\theta,\phi)&=&\frac{3\Gamma\sin^2{\theta}}{\pi N^2}\sum_{mn}\frac{B_m(\theta,\phi)B^*_n(\theta,\phi)}{D_m+D_n^*}\\\nonumber
  &&\times \sum_{jk}C^{(p)}_{jk} e^{-i\mathbf{k}_\mathrm{L}\cdot\mathbf{r}_k} e^{i\phi_k(n-1)} e^{i\mathbf{k}_\mathrm{L}\cdot\mathbf{r}_j} e^{-i\phi_j(m-1)},
\end{eqnarray*}
where we have abbreviated $C^{(p)}_{jk}=\sum_{j'}\psi_{jj'}^{(p)}{\psi_{j'k}^{(p)}}^*$,
and where the expression of $B_n(\theta,\phi)$ is given in Eq. (\ref{eqn:B}). A comparison of the numerical results indeed shows that the intensity profiles of the photon pair $\left|\Phi_{2_1}\right>$ and of the single photon $\left|\Phi_1\right>$ are virtually identical for every parameter regime.

Finally, to complete the analysis of the photon pairs, we calculate the density-density correlations. They are defined as the probability of detecting a photon in the solid angle $\Omega_{\mathbf{q}'}$ upon detecting a photon in solid angle $\Omega_{\mathbf{q}}$ normalized to the probability of uncorrelated detection, i.e.,
\begin{equation*}
  g_2(\Omega_{\mathbf{q}},\Omega_{\mathbf{q}'})=\frac{G(\Omega_{\mathbf{q}},\Omega_{\mathbf{q}'})}{I(\Omega_{\mathbf{q}})I(\Omega_{\mathbf{q}'})}-1,
\end{equation*}
where $G(\Omega_{\mathbf{q}},\Omega_{\mathbf{q}'})\propto\int_0^\infty \omega^2d\omega\int_0^\infty {\omega'}^2d\omega'\sum_{\nu\nu'}\left<n_{\mathbf{q}\nu}n_{\mathbf{q}'\nu'}\right>$.
This correlation function yields $g_2(\Omega_{\mathbf{q}},\Omega_{\mathbf{q}'})=0$ if the two photons that are detected at the angular coordinates $\Omega_{\mathbf{q}}$ and $\Omega_{\mathbf{q}'}$, respectively, are uncorrelated, and $g_2(\Omega_{\mathbf{q}},\Omega_{\mathbf{q}'})>0$ ($<0$) for correlation (anticorrelation) between the two photons.

We study the correlations as follows: First, the angular distribution of photons is obtained. We then assume that the first photon is detected in one of the maxima of this distribution. Subsequently, we calculate and plot the correlations as a function of the angular coordinates of the second photon.

\begin{figure}
\center
\includegraphics[width=6cm]{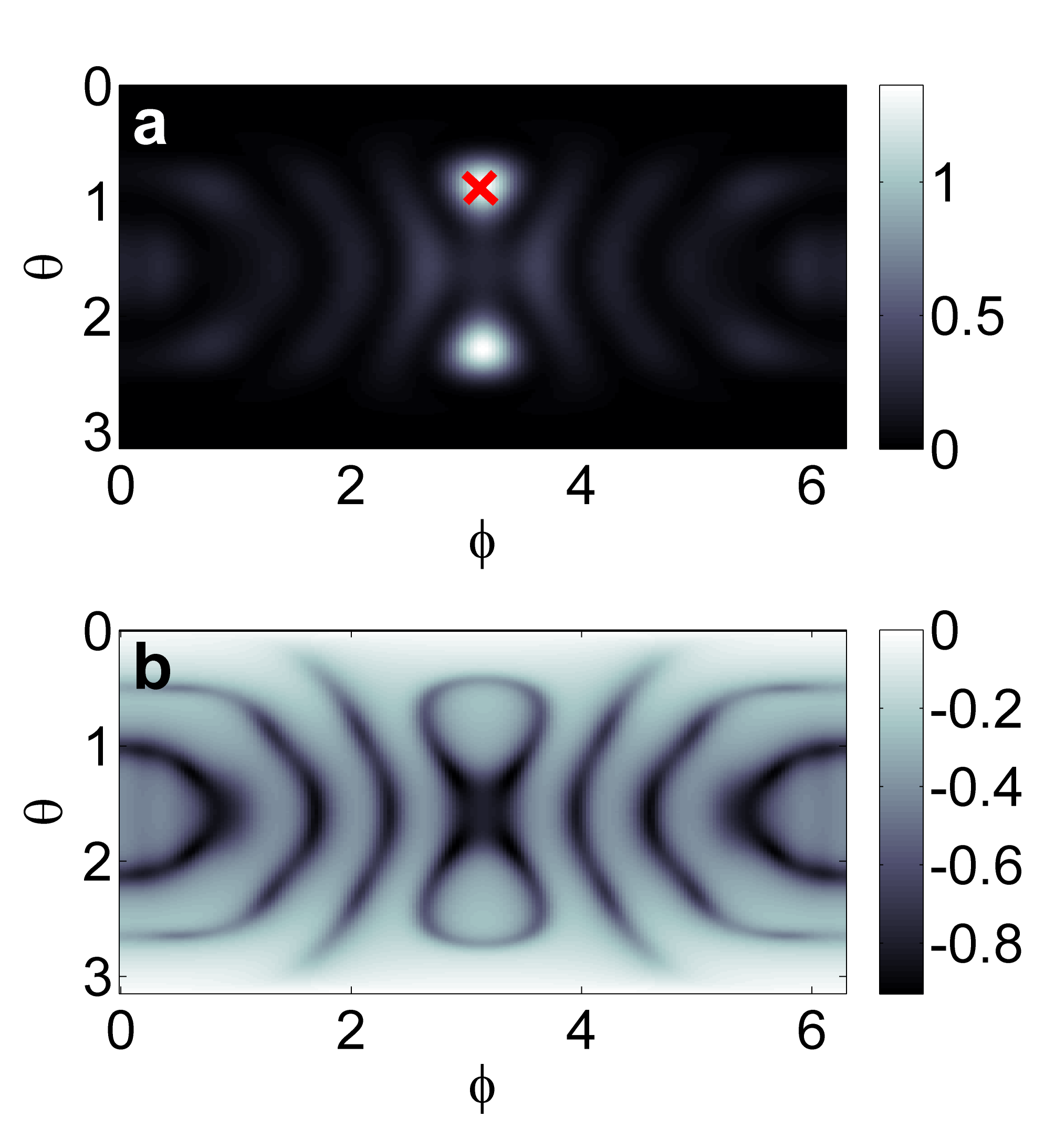}
\caption{\textbf{a}: Angular intensity distribution and \textbf{b}: density-density correlation function for $N=15$ and $a=\lambda_\mathrm{L}/2$ of the two-photon state $\left|\Phi_{2_1}\right>$. The correlations are calculated fixing one of the angles to the maximum of the intensity marked by the red cross, i.e., $\left(\theta_\mathrm{max},\phi_\mathrm{max}\right)=\left(0.83,\pi\right)$ (note that the angle of incidence of the laser is $\left(\theta_\mathrm{L},\phi_\mathrm{L}\right)=\left(\pi/4,\pi\right)$). There is anticorrelation for all angles.} \label{fig:correlations1}
\end{figure}

Let us commence by studying the state $\left|\Phi_{2_1}\right>$ which is shown in Fig. \ref{fig:correlations1}. We observe that the correlation function is negative (i.e., anticorrelation of the photons) for all angles $\Omega_{\mathbf{q}}\neq\Omega_{\mathbf{q}'}$. This in turn means that the probability of detecting the two photons in the same direction is very high. This confirms our previous observation, i.e., that to a good degree of approximation both photons are emitted into the same state. The particular features of the angular intensity and the density-density correlation function of this photonic state for $N=15$ are shown in Fig. \ref{fig:correlations1}a and b where the angle of the first photon detection is indicated in the intensity profile by a red cross.

\begin{figure}
\center
\includegraphics[width=6cm]{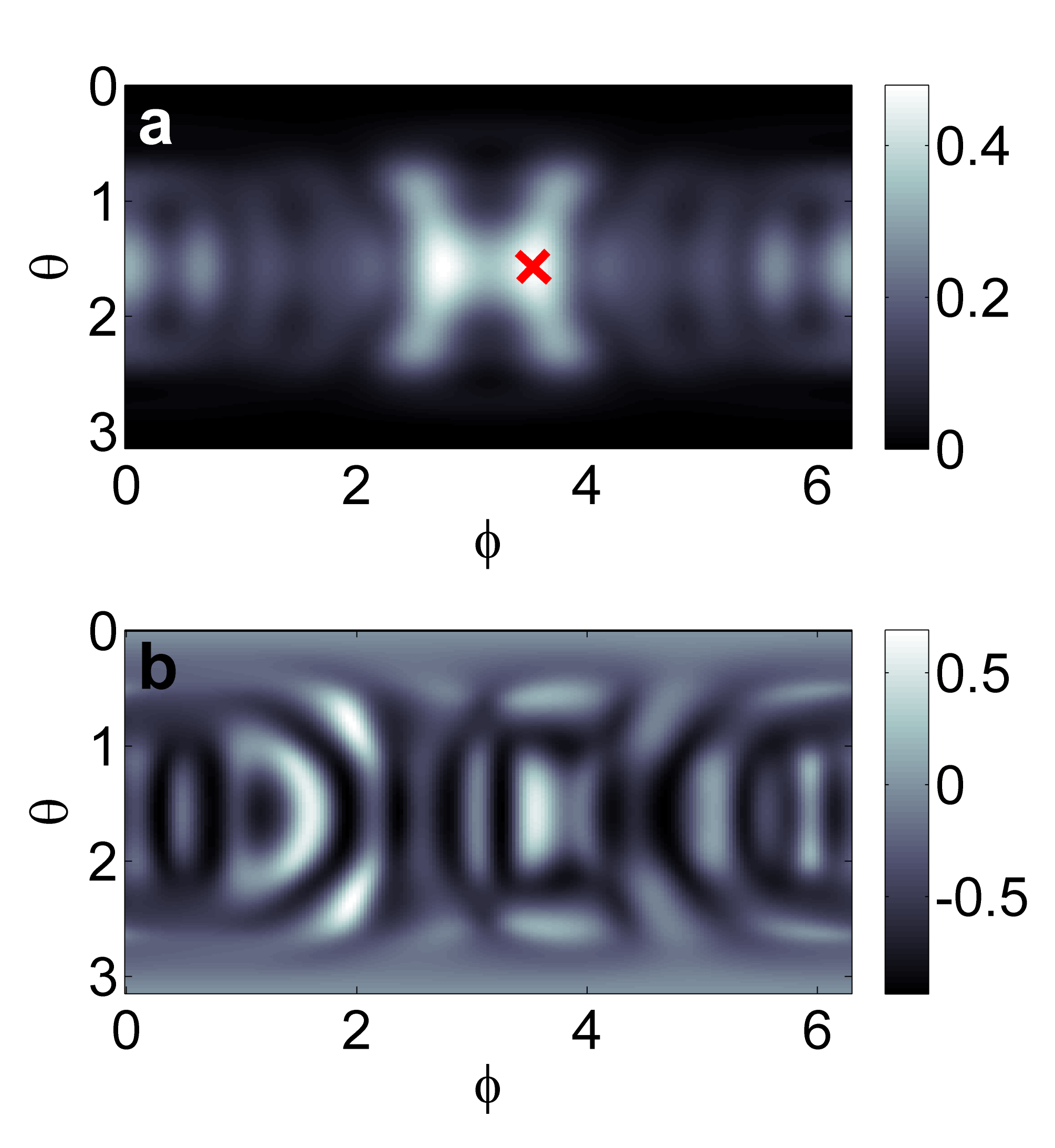}
\caption{\textbf{a}: Angular intensity distribution and \textbf{b}: density-density correlation function for $N=15$ and $a=\lambda_\mathrm{L}/2$ of the two-photon state $\left|\Phi_{2_3}\right>$. The correlations are calculated fixing one of the angles to the maximum of the intensity marked by the red cross, i.e., $\left(\theta_\mathrm{max},\phi_\mathrm{max}\right)=\left(\pi/2,3.52\right)$ (note that the angle of incidence of the laser is $\left(\theta_\mathrm{L},\phi_\mathrm{L}\right)=\left(\pi/4,\pi\right)$).} \label{fig:correlations2}
\end{figure}

Let us now investigate the correlations for other photonic states with $p\neq 1$. As discussed before, in general the pairs of photons resulting from the atom-photon mapping are not emitted any longer into a product state but into a non-classical entangled one. Hence, new features of the correlations are expected to appear. As an example, in Fig. \ref{fig:correlations2}a and b we show the angular distribution of photons $I(\theta,\phi)$ and the density-density correlation function $g_2(\Omega_{\mathbf{q}},\Omega_{\mathbf{q}'})$ for $p=3$ and $N=15$. In this case, the correlation pattern is rather complex. However, unlike for the $p=1$ case one observes some very pronounced highly correlated peaks in the half-space $0\leq\phi<\pi$. Note that the maximum probability for the detection of the first photon (maximum of the intensity) was in the opposite half-space $\pi<\phi\leq 2\pi$.

\section{Conclusions and outlook} \label{sec:conclusion_outlook}
In this work, we have first reviewed the problem of outcoupling non-classical light from a coherently prepared atomic ensemble. We have explained how for large enough times this process provides a direct mapping between atomic excitations and photonic states. Subsequently we have applied this atom-photon mapping to a set of many-body entangled states that can be created on a ring-shaped lattice.

Our analysis of the single photon states that are obtained via the excitation of an atomic spin wave in the ring has revealed two results: First, we have observed that, when the wavelength of the driving laser is smaller than the extension of the system, the photon is emitted into a superposition state of two different directions. These directions can be tuned by varying the incident angle of the laser with respect to the ring. Second, in the regime where the interatomic separation is of the order of the wavelength $\lambda_\mathrm{L}$, we have shown that it is possible to produce hollow photons with well-defined zero orbital angular momentum. This is achieved even for small atomic ensembles ($N\sim10$) due to collective effects in the photon emission.

In addition, we have studied photon pairs that are created from doubly excited entangled states which can be prepared in the ring lattice. During our analysis of the angular density-density correlations of these photon pairs we found that the doubly excited atomic states are in general mapped into entangled photons with non-trivial correlation properties.

The feasibility to create two-dimensional arrays of optical traps which is necessary for the experiment implementation of our scheme has already been experimentally demonstrated \cite{Kruse10,Bergamini04}. These trap arrays with site separations on the order of a micrometer and single-site addressability can be used to create experimentally a ring lattice and other complex 2D configurations such as triangular or hexagonal lattices. Studying the creation of many-body entangled atomic states along the lines of Refs. \cite{Olmos09,Olmos09-3} in these setups is thus an interesting and exciting future direction - in particular because other geometries are expected to lead to new features of the photonic states. Moreover, the finite strength of the confinement of the atoms in conjunction with a finite temperature will lead to an uncertainty of the atomic positions $\textbf{r}_\gamma$. This will produce disorder that affects the preparation of the entangled atomic resource states \cite{Olmos09-3} as well as the intensity distribution and correlation properties of the emitted photons \cite{Porras08}. This will be explored in detail  in a future work.

The authors acknowledge funding by EPSRC.

\end{document}